\documentclass[twocolumn,showpacs,preprintnumbers,amsmath,amssymb]{revtex4}
\usepackage[T1]{fontenc}
\usepackage[english]{babel}
\usepackage[dvips]{graphics}

\newcommand{\beq}{\begin{equation}}
\newcommand{\eeq}{\end{equation}}
\newcommand{\bma}{\begin{math}}
\newcommand{\ema}{\end{math}}
\newcommand{\beqa}{\begin{eqnarray}}
\newcommand{\eeqa}{\end{eqnarray}}

\def\half{\frac{1}{2}}
\def\opone{\le\textbf{}\textbf{}avevmode\hbox{\small1\kern-3.8pt\normalsize1}}

\begin{document}

\title{Half-Filled Lowest Landau Level on a Thin Torus}

\author{Emil J. Bergholtz}

\email{ejb@physto.se}

\author{Anders Karlhede}

\email{ak@physto.se}

\affiliation{Department of Physics,
Stockholm University \\ AlbaNova University Center\\ SE-106 91 Stockholm,
Sweden}

\date{\today}

\begin{abstract} We solve a model that describes an interacting electron gas in the half-filled lowest Landau level on
a thin torus, with radius of the order of the magnetic length. The low energy sector consists  of non-interacting, 
one-dimensional, neutral fermions.  The ground state, which is homogeneous, is the Fermi sea obtained by filling 
the negative energy states and the excited states are gapless neutral excitations out of this one-dimensional sea. Although the limit considered is extreme, the solution has a striking resemblance to the composite fermion 
description of the bulk $\nu=1/2$ state---the ground state is homogeneous and the excitations are neutral and gapless.  
This suggests a one-dimensional Luttinger liquid description, with possible observable effects in transport experiments, of the bulk state where it develops continuously from the state on a thin 
torus as the radius increases.   \end{abstract}


\maketitle

It was observed by Jiang {\em et al} in 1989 that the
quantum Hall (QH) system has a metallic behavior at filling fraction $\nu=1/2$ ---$\sigma_{xx}$ is finite and sample
dependent as $T\rightarrow 0$, whereas $\sigma_{xy}$ is unquantized and approximately equal to
its classical value  \cite{jiang}. Early experiments  also showed that there is a large density of low energy states
\cite{willett90}, but no nonlocal transport  \cite{wang}. 

The metallic $\nu=1/2$ state was successfully described by Halperin, Lee and Read \cite{hlr}, who
introduced a  mean field theory where the external magnetic field is cancelled by a smeared out
statistical field,  resulting in composite fermions \cite{jain} moving
in a zero field---{\em ie} in a two
dimensional free fermion gas with a Fermi surface. This picture, as well as Jain's general concept of composite
fermions, was spectacularly confirmed by surface acoustic wave experiments performed by Willett {\em et al}
\cite{willett}, and by ballistic transport \cite{ballistic},
which showed that particles move in a reduced effective magnetic field (or, alternatively,
have a reduced charge) which approaches zero as $\nu \rightarrow 1/2$.

Rezayi and Read \cite{rr} proposed a microscopic wave function for the $\nu=1/2$ state, which
agrees very well with exact results for small systems. The theory for the $\nu=1/2$ state was later
further developed by several groups \cite{several}, and a description in terms of neutral dipoles was proposed.

In spite of the impressive agreement between theory and experiment there are, in our opinion, important questions regarding the physics in 
the lowest Landau level that remain to be answered.  There is no real understanding of why the strongly correlated electron system 
in the lowest Landau level, at various filling fractions, becomes a system of weakly interacting composite fermions---no controlled 
microscopic derivation of the mean field theory at $\nu=1/2$ or, for that matter, of the composite fermion descriptions at other filling fractions exists. 

Here we study the interacting electron gas in the lowest Landau level at $\nu=1/2$ on a thin torus. We obtain an 
exact solution for a particular short-range electron-electron interaction that is relevant for a torus with circumference, $L_1$, 
of the order of the magnetic length. 
The low-energy sector consists of  free neutral one-dimensional fermions. Expressed in terms of the original electrons, these "composite 
fermions" are nearest neighbor electron-hole pairs, excitons, with a hard-core constraint. This thus provides a dipole picture of 
the $\nu=1/2$ state.  The
ground state is a homogeneous Fermi sea of the neutral fermions, which supports gapless neutral excitations.
\begin{figure}[h!]
\begin{center}
\resizebox{!}{12mm}{\includegraphics{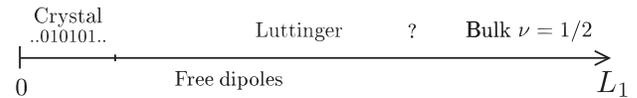}}
\end{center}\caption{\textit{{\small Phase diagram for $\nu=1/2$ on a torus, where one circumference, $L_1$, varies while 
the other is infinite. For a thin torus, $L_1 \sim 5$ magnetic lengths, the $\nu=1/2$ system is that of noninteracting neutral 
one-dimensional fermions (dipoles). 
The question mark indicates whether the state develops continuously into the bulk $\nu=1/2$ state or whether there is a 
phase transition. Based on the similarities of the state at short $L_1$ and the bulk state, such as a homogeneous ground state and 
neutral gapless excitations, we conjecture that there is no phase transition. This suggests a one-dimensional description 
of the bulk $\nu=1/2$ state as a Luttinger liquid rather than as a free two-dimensional Fermi gas.
(For very short $L_1$ a crystalline state determined by electrostatics alone is obtained.)}}}
\label{qh1}
\end{figure}

The low-energy sector has many features in common with the bulk
$\nu=1/2$ state, such as a homogeneous ground state and gapless
neutral excitations, and
we conjecture that it develops continuously into the bulk state as $L_1 \rightarrow \infty$, rather than
being separated from this by a phase transition, see Fig. 1. This suggests a one-dimensional description
of the  bulk $\nu=1/2$ state  as a 
Luttinger liquid rather than as a two-dimensional Fermi gas \cite{Lee}. 

We start by mapping the problem onto a one-dimensional lattice model.
Following Haldane and Rezayi we  consider an electron confined to the lowest Landau level
on a torus with lengths $L_1, L_2$ in the $x$ and $y$ directions, respectively \cite{Haldane85}.
In Landau gauge, $\vec A= - By\hat x$, the magnetic translation operators become
\beqa
\label{translations}
T_1=  e^{(L_1/N_s)\partial_x} \  \  , \ \ \  T_2=e^{(L_2/N_s)(\partial_y+ix)} \  \ ,
\eeqa
where $N_s$ is the number of flux quanta through the surface of the torus. (The magnetic length is set to one,
$\ell = \sqrt{\hbar c/eB}=1$.) These operators commute
with the Hamiltonian for the  charged particle coupled to $\vec A$.
Wave functions are required to be periodic up to a phase,
$T_\alpha^{N_s} \Psi = e^{i \phi_\alpha} \Psi$, $\alpha =1,2$, leading to $L_1L_2= 2\pi N_s$ and
$T_1 T_2 = e^{2\pi i/N_s}T_2 T_1$. With $\psi_0 = \sum_{n} e^{inL_2x}e^{-(y+nL_2)^2/2}$, we 
obtain the $T_1$ eigenstates  $\psi_m=T_2^m \psi_0$,  $T_1 \psi_m= e^{i 2\pi m/N_s}\psi_m$,
$m=0,1,... N_s-1$. The states $\psi_m$  span  the lowest
Landau level  and are located along the lines $y=-2\pi m/L_1$. Thus we have obtained the
mapping onto a one-dimensional lattice model, where $m$ numbers the sites and the lattice  constant is
$2\pi/L_1$.

Assuming translation invariance, the electron-electron interaction Hamiltonian becomes
\beqa
\label{ham}
H_{ee} =\sum_n \sum_{k > m} V_{km}c^\dagger_{n+m}c^\dagger_{n+k}c_{n+m+k}c_n \ \ ,
\eeqa
where $c^\dagger_m$ creates an electron in state $\psi_m$ and $V_{km}=V_{k,-m}\ge 0$.  To understand
the physics of the interaction it is useful to divide $H_{ee}$ into two parts:  $V_{p0}$, the electrostatic
repulsion between two electrons  separated $p$ lattice constants, and $V_{m+p,m}$, the amplitude for
two particles separated a distance $p$ to
hop symmetrically to a separation  $p+2m$ and vice versa.
For a short-range real space electron-electron
interaction of the form  $V(\vec r) = \nabla^2 \delta(\vec r)$ 
one  finds $V_{km}=(k^2-m^2)e^{-2 \pi^2 (k^2+m^2)/L_1^2}$ \cite{cylinder}. When the torus becomes thin, {\it ie}
when $L_1$ decreases, the distance $2\pi/L_1$ between the single particle states increases, hence
fewer terms in (\ref{ham}) contribute. For the $\nabla^2 \delta$ interaction one finds that the range
of the interaction is of the order of six magnetic lengths.

We consider the electron gas at filling fraction $\nu=1/k$, where $k$ is an  integer,  and assume that
the number of electrons $\nu N_s$ is an integer. The many particle states can be chosen
as $T_1$ eigenstates with momentum $\kappa$ mod($N_s$) (in units of $2\pi/N_s$). $T_2$ translates the system
in the $y$-direction and changes $\kappa$ by $\nu N_s$---{\it ie} by the number of particles. Since $T_2$
commutes with the Hamiltonian, all energy eigenstates
are $k$-fold degenerate \cite{Haldane85}. ($T_2^k$ preserves $\kappa$ mod($N_s$) and hence can be diagonalized
along with $T_1$.)

At a fixed filling fraction, the electrostatic repulsion strives to keep the particles apart,
whereas the hopping terms favour  maximally hoppable states. To find the low energy states is
in general a very  complicated problem. However,  at $\nu=1/2$, the short range electrostatic and hopping terms 
cooperate, leading to a simple low-energy sector for the thin torus.

We truncate the interaction in (\ref{ham}) and keep only the two shortest range electrostatic terms and the shortest range hopping term
\beqa
\label{hamtrunc}
H=
&\sum_{n} [V_{10} \hat n_n \hat n_{n+1} +V_{20} \hat n_n \hat n_{n+2}  \nonumber \\
&-  V_{21} ( c^\dagger_{n}c_{n+1}c_{n+2}c^\dagger_{n+3} + {\rm H.c.})] \ \  ,
\eeqa
where $\hat n_k=c_k^\dagger c_k$.  This provides  a good approximation of the interaction
on a thin torus as discussed below \cite{parameters}.

Before giving the details of our analysis we outline the logic of the identification of the low-energy sector of the 
Hamiltonian (\ref{hamtrunc}) at $\nu=1/2$.  The crucial  part in (\ref{hamtrunc}) is the hopping term 
$V_{21}$. We define a subspace  ${\cal H}^\prime$ of the full Hilbert space by requiring each pair of sites $(2p-1,2p)$ to 
have charge one. Acting with $T_2$ gives an equivalent  grouping of the
sites  $(2p,2p+1)$ instead---and a corresponding subspace
 $\cal H ^\prime _T$ \cite{twostates}.   ${\cal H}^\prime$ (and $\cal H ^\prime _T$) is 
the low-energy sector under fairly general conditions since it contains the maximally hoppable state 
$|100110011001....\rangle$---which turns out to be the seed for the ground state and is also the lowest energy state for 
the $V_{20}$-term---and it has a low electrostatic energy by construction.   
$H$ preserves the subspace $\cal H^\prime$ and the hopping term can be exactly diagonalized in this space giving non-interacting neutral fermions.
The ground state is the one-dimensional Fermi line obtained by filling the negative energy states, and 
the excitations are gapless excitations out of this Fermi sea. The electrostatic terms in (\ref{hamtrunc}) are less crucial. At  $V_{10}=2 V_{20}$ 
all states in $\cal H^\prime$ have the lowest possible electrostatic energy and we argue  perturbatively that  $\cal H^\prime$ is the low energy sector.  
However, we expect this to be true under more  general conditions. 

We now present our analysis for the truncated Hamiltonian (\ref{hamtrunc}) at $\nu=1/2$ and $V_{10}=2 V_{20}\equiv 2\alpha$ .
The electrostatic part,  $H|_{V_{21}=0}$, then has the eigenstates  $|n_1n_2...n_{N_s}\rangle$,
where $n_i=0,1$ and $|1\rangle = c^\dagger|0\rangle$, with energies
\beqa
\label{elstat}
E_0=\alpha {\Big ( }\frac {N_s} 2 + n_{111} + n_{000}{\Big )}\ \ .
\eeqa
Here, $n_{111} \, (n_{000})$ is the number of 3-strings, {\em ie} strings consisting of three nearby electrons
(holes) in $n_1n_2...n_{N_s}$ (a string
of length $k\ge 3$ is counted as $k-2$ strings and periodic boundary conditions are assumed).
Thus there is a degenerate ground state manifold ${\cal H}_0$ consisting of all states where at most two electrons or two holes are
next to each other. Note that ${\cal H}^\prime \subset {\cal H}_0$. The excitations are 3-strings  of either electrons or
holes and each 3-string has energy $\alpha$. The statement about ${\cal H}_0$ follows by induction if the states without 3-strings for
$N_s$ sites are
constructed by inserting an electron and a hole in  $N_s-2$ states without 3-strings. The energy of a 3-string follows by considering the change
in energy of one or several 3-strings when moving one constituent.

To diagonalize $H$ in $\cal H ^\prime$, we proceed as follows.
There are two possible states for a pair of sites in $\cal H ^\prime$: $|\downarrow \rangle \equiv |01\rangle$ and
$|\uparrow \rangle \equiv |10\rangle$ and it is natural to introduce the spin raising operator
$s^+_p=c^\dagger_{2p-1}c_{2p}$, $|\uparrow\rangle =s^+ |\downarrow\rangle$.
On states in $\cal H ^\prime$, $s^+, \, s^-=(s^+)^{\dagger}$ describe hard core bosons---they
commute on different sites but anticommute on the same site and $H$ is
the nearest neighbor spin-1/2 $XY$ chain. Expressing the
bosons in terms of fermions $d$  using the Jordan-Wigner transformation, $s^-_p=K_p d_p$, where
$K_p=e^{i\pi \sum_{j=1}^{p-1}d^\dagger_j  d_j}$, the
Hamiltonian (\ref{hamtrunc}) is simply that of free fermions, $H=\frac {\alpha N_s} 2+V_{21} [\sum_{p=1}^{N_s/2-1} 
d^\dagger_{p+1} d_{p} +d^\dagger_{1} K_{N_s/2}d_{N_s/2}+H.c.]$,
when restricted to $\cal H ^\prime$ \cite{sign}. 
Thus, after a Fourier transformation, the ground state is obtained by filling all the negative energy states. This state has energy $E=\frac \alpha 2 - \frac {V_{21}} {\pi}$ 
per site (if $N_s\rightarrow \infty$) and supports neutral gapless excitations. 
One readily finds that $\langle c^\dagger_m c_n\rangle=\half \delta_{mn}$, and hence the state is homogeneous.

This solves the problem in  $\cal H ^\prime $ and, by action of $T_2$,  in $\cal H ^\prime _T$.  It
remains to consider
the states in the ground state manifold ${\cal H}_0$ that are in neither of these subspaces. We will now argue
that these are separated from the ground state by a gap of order $V_{21}$ generated by the hopping term in $H$. Intuitively, this
makes sense since $\cal H ^\prime$ contains the maximally hoppable state $|01100110011....110\rangle$. Note that whereas
 $\cal H ^\prime $ is invariant under $H$, other states in ${\cal H}_0$   may mix with states not in the ground state manifold.
Our procedure will be to simply diagonalize $H$ in the ground state manifold ${\cal H}_0$.

To describe a  general state in ${\cal H}_0$, we introduce the notation $|a\rangle \equiv |00\rangle$ and $|b\rangle \equiv |11\rangle$, along
with $|\downarrow \rangle$ and $|\uparrow \rangle$, for the states on sites $(2p-1,2p)$. $H$ contains the hopping terms:  
$\uparrow\downarrow  \leftrightarrow  \downarrow \uparrow$,
$\downarrow a \uparrow  \leftrightarrow  aba$,  $\uparrow b \downarrow \leftrightarrow bab$, $\uparrow ba \leftrightarrow ba \uparrow$ and
$\downarrow ab \leftrightarrow ab \downarrow$---all with strength $V_{21}$.
A general state is uniquely described by a string of $\uparrow, \downarrow, a$ and $b$ and can be characterized by the 
number $d$ of alternating $\cal H^\prime$ and $\cal H ^\prime _T$ domains it is built up of.
Any pair $\uparrow \downarrow$ or $ \downarrow \uparrow$ belongs to $\cal H^\prime$ and any $a$ or $b$ belongs to  $\cal H ^\prime _T$---a 
polarized string $\uparrow \uparrow \uparrow  \uparrow....$ or $\downarrow \downarrow \downarrow \downarrow....$ can however belong to either domain.
This implies that there is a domain wall in between $\uparrow \downarrow$ (or $ \downarrow \uparrow$) and  $a $ (or $b$)---counting 
the number of domain walls gives $d$ for a general state.
 
A state is a $\nu=1/2$ state if it has an equal number of $a'$s and $b'$s and belongs to 
${\cal H}_0$ if  it does {\it not} contain any of
the nearest neighbor combinations $(aa), \ (bb), \ (a\downarrow), \ \newline (b\uparrow), \, (\uparrow a)$ or $(\downarrow b)$. 
It is straightforward to show that $d$ is preserved by $H$. The states in  $\cal H ^\prime $ and  $\cal H ^\prime _T$ are the 
$d=1$ states. In the $d=2$ sector,
we consider first the states with one $a$ and one $b$ next to each other, $ab$ or $ba$,  in a string of spins. These $d=2$ states are mapped into
each other under $H$. To be in  ${\cal H}_0$, $ab$ must enter as $X\equiv \downarrow ab \downarrow$. Under $H$, this hops just like  $\uparrow$:
$X\downarrow \leftrightarrow \downarrow X$ with matrix element $V_{21}$. Thus the problem is equivalent to the $\cal H ^\prime $ problem
with $N_s - 6$ sites and one finds that these states are separated from the ground state in
$\cal H ^\prime $ by a gap of order $V_{21}$. By considering how hoppable the states are we expect the states just considered to be the lowest energy states
in the $d=2$ sector. We have verified this by exact diagonalization in ${\cal H}_0$ of up to $N_s=18$ sites.
The $d \ge 3$ states contain more domain walls and it is easy to see that the hopping becomes more restricted---thus we expect them to have higher energy.
We have verified this numerically for $N_s \le 18$. Thus we conclude that $\cal H ^\prime $ and $\cal H ^\prime _T$ give the
low-energy sector of the theory---the remaining states in ${\cal H}_0$ are separated from the ground state by a gap of order $V_{21}$.

We now consider the stability  of the solution we have obtained for $H$ in  (\ref{hamtrunc}) when $V_{10}=2V_{20}$ 
and investigate whether it  describes the $\nu=1/2$ state on a thin torus for a range of $L_1$. 
We first note that for the real space short-range interaction  $V(\vec r) = \nabla^2 \delta (\vec r)$,  
$V_{10}=2V_{20}=2\alpha$ corresponds to
$L_1=2\pi/\sqrt{2\ln 2}=5.3$. The hopping term included in (\ref{hamtrunc}) is then $V_{21}=\frac 3 8 \alpha$,  whereas  the leading ignored terms
are small: $V_{30}=\frac 9 {128} \alpha$ and $V_{31}=\frac 1 {32} \alpha$. This is close to the solvable point. 

We  have performed density matrix renormalization group (DMRG) \cite{White} studies on a thin cylinder with the Hamiltonian (\ref{ham}) 
and $V(\vec r) = \nabla^2 \delta (\vec r)$ including interactions that extend over up to six lattice constants  \cite{bergholtz03}. We find a 
ground state that is homogeneous to very high accuracy and strong indications of gapless excitations in the region 
around $L_1=2\pi/\sqrt{2\ln 2}$ that we have investigated ($4\lesssim L_1\lesssim 8$). 
(When $L_1$ is even smaller, the ground state is a crystal 
$|\uparrow\uparrow\uparrow\uparrow\uparrow\uparrow....\uparrow\rangle $---the lowest energy state for the shortest range 
electrostatic term $V_{10}$.)  

The low energy sector at the solvable point ($H$ in (\ref{hamtrunc}) with $V_{10}=2V_{20}$) is contained in the spin-1/2 
Hilbert space $\cal H^\prime$. The states not in $\cal H^\prime$  are separated from the low energy states by a gap.
Small perturbations of the Hamiltonian around the solvable point lead, in perturbation theory,  to an effective spin-1/2 Hamiltonian in $\cal H^\prime$. 
The generated terms are spin operators of quadratic and higher order. They all have small coefficients
since there is a gap to states not in $\cal H^\prime$ (the matrix elements for transitions to states in  $\cal H^\prime _T$ vanish). 
All  terms are irrelevant in the sense of the renormalization group, except for  $s^z_i s^z_{i+n}$ which 
makes the noninteracting fermion theory develop into a Luttinger liquid with interaction parameter $K\neq 1$ (at the solvable point $K=1$), see {\it eg}  \cite{schulz}. 

To obtain the general effective Hamiltonian explicitly is non-trivial. However, the electrostatic terms, $V_{m0}$, preserve $\cal H^\prime$ and simply become  $\sum_{i,n} [(2V_{2n,0}-V_{2n-1,0}-V_{2n+1,0})s^z_is^z_{i+n}]$. The hopping terms, $V_{mn}$,  will, in general,  contribute  in second order perturbation theory.  

Based on the renormalization group argument, and supported by the DMRG calculations, we conclude that  the $\nu=1/2$ system on a thin torus is a 
Luttinger liquid for a finite range of $L_1$ and that the generation of $s^z_i s^z_{i+n}$ terms indicates that the interaction parameter that determines the decay of correlation functions is shifted from its value at the solvable point. 

When $L_1$ increases further, there is either a phase transition or the state develops continuously into the bulk $\nu=1/2$ state. We conjecture that the latter is the case. 
The main support for this comes from the striking similarities of the 
low-energy sector on the thin torus and the composite fermion description of the bulk state, most notably 
the homogeneous ground state and the gapless neutral excitations.  Furthermore, we note that the
reduction of the Hilbert space to ${\cal H}^\prime$ by itself implies that the charge on average is homogeneous---this, or some suitable generalization thereof,
is likely to be a good approximation also when $L_1$ increases and longer range interactions come into play. 

Further support for our conjecture comes from  considering the Laughlin filling
fraction $\nu=1/3$ \cite{Laughlin}. Rezayi and Haldane  have shown that  the Laughlin state is the $\nu=1/3$ ground state also on a thin cylinder 
and  that it develops continuously from a charge density wave state into the homogeneous Laughlin state
as $L_1\rightarrow \infty$ \cite{Haldane94}. Our DMRG calculations agree with this---we find, using (\ref{ham}), for a range of $L_1$ a 
charge density wave state in quantitative agreement with that of  Rezayi and Haldane. Thus the $\nu=1/3$ ground state of the short range
Hamiltonian develops continuously into the homogeneous Laughlin state
as $L_1\rightarrow \infty$. This lends some support for our conjecture that the  $\nu=1/2$ state also develops adiabatically. However,
in this case there is no gap and the  issue is more delicate. The argument would be strengthened if the picture of the $\nu=1/2$ state given above 
could be shown to generalize to the $\nu=1/3$ state on the thin torus, in which  case it should be
relevant also for the bulk $\nu=1/3$ state. The mapping of the low energy sector at $\nu=1/2$ onto an $s=1/2$
$XY$-spin chain, would then presumably generalize into a mapping of  $\nu=1/3$
onto an $s=1$ chain. In passing, we note that this suggests that the Haldane conjecture for the gaps in spin chains \cite{haldane} might apply to the 
two-dimensional electron gas in a strong magnetic field.

The thin torus, or cylinder for that matter, with a magnetic field perpendicular to its surface is probably not experimentally accessible. Thus, 
the experimental consequences of the results in this Letter presumably depend on whether the results are applicable, {\it mutatis mutandis},  
to the bulk case as we conjecture. Our conjecture implies that the $\nu=1/2$ state is a one-dimensional Luttinger liquid rather than a two-dimensional
Fermi theory. We predict that this leads to observable effects in the bulk $\nu=1/2$ system, such as 
non-linear $I-V$ characteristics determined by the Luttinger liquid interaction parameter. 

If our conjecture is correct, then it should  be possible  to understand the experimental results that are successfully explained by 
the standard composite fermion theory \cite{hlr}, such as the ballistic transport 
and the surface acoustic wave results. We note that the appearance of low-energy excitations 
that are neutral, and hence do not couple to the magnetic field, is  consistent with the ballistic transport results.

We thank Thors Hans Hansson and Jon Magne Leinaas for valuable discussions. Anders Karlhede was supported by the Swedish Research Council.

\end{document}